\documentclass[aps,prl,reprint,amsmath,amssymb,showpacs,superscriptaddress]{revtex4-1}
\usepackage{CJK}
\usepackage{graphicx}
\usepackage{dcolumn}
\usepackage{bm}
\usepackage{color}
\usepackage{hyperref}

\allowdisplaybreaks[4]

\begin{document}

\title{Relativistic effects and three-body interactions in atomic nuclei}
\author{Y. L. Yang}
\affiliation{State Key Laboratory of Nuclear Physics and Technology, School of Physics, Peking University, Beijing 100871, China}

\author{P. W. Zhao}
\email{pwzhao@pku.edu.cn}
\affiliation{State Key Laboratory of Nuclear Physics and Technology, School of Physics, Peking University, Beijing 100871, China}

\begin{abstract}
Based on the leading-order covariant pionless effective field theory, a relativistic nuclear Hamiltonian is derived and solved using the variational Monte Carlo approach for $A\le 4$ nuclei by representing the nuclear many-body wave functions with a symmetry-based artificial neural network.
It is found that the relativistic effects rescue the renormalizability of the theory, and overcome the energy collapse problem for $^3$H and $^4$He without promoting a repulsive three-nucleon interaction to leading order as in nonrelativistic calculations.
Nevertheless, to exactly reproduce the experimental ground-state energies, a three-nucleon interaction is needed and its interplay with the relativistic effects plays a crucial role.
The strongly repulsive relativistic effects suppress the energy contribution given by the three-nucleon interactions, so a strong strength for the three-nucleon interaction could be required to reproduce the experimental energies.
These results shed light on a consistent understanding of relativistic effects and three-body interactions in atomic nuclei.
\end{abstract}

\maketitle

{\it Introduction.}---One of the main challenges of nuclear physics is to
explain the spectra, structure, and reactions of nuclei starting from the underlying nuclear interactions, which are usually constructed to reproduce few-body observables and then used as inputs for \textit{ab initio} calculations to predict properties of finite nuclei and nuclear matter.
Already in the 1980s, accurate three-body calculations showed that contemporary two-body nucleon-nucleon ($NN$) interactions fit to scattering data cannot provide enough binding for three-body nuclei, ${^3}$H and ${^3}$He~\cite{Friar1984Annu.Rev.Nucl.Part.Sci.403}.
Subsequently, the importance of three-nucleon ($3N$) interactions was confirmed by the quantum Monte Carlo (QMC) studies for the ground and low-lying excited states of light nuclei up to $A=16$~\cite{Carlson2015Rev.Mod.Phys.1067,Lonardoni2018Phys.Rev.Lett.122502}.
Nowadays, it has been a usual consensus that apart from the $NN$ interactions, the $3N$ interactions should also be crucial to achieve a complete description of nuclear systems.

While many \textit{ab initio} studies are performed nonrelativistically, there are strong evidences showing that relativistic effects could play an important role in constructing nuclear interactions~\cite{Ren2018Chin.Phys.C14103,Lu2022Phys.Rev.Lett.142002} and in \textit{ab initio} calculations of nuclear matter~\cite{Brockmann1984Phys.Lett.B283, Brockmann1990Phys.Rev.C1965} and finite nuclei~\cite{Stadler1997Phys.Rev.Lett.26,Forest1999Phys.Rev.C014002,Shen2019Prog.Part.Nucl.Phys.103713}.
There are essentially two different frameworks to estimate the relativistic effects on nuclear binding.
One is linked to the manifestly covariant quantum field theories~\cite{Rupp1992Phys.Rev.C2133,Stadler1997Phys.Rev.Lett.26,Shen2019Prog.Part.Nucl.Phys.103713}, and the other one is based on the relativistic quantum mechanics with a relativistic Hamiltonian ~\cite{Bakamjian1953Phys.Rev.1300, Krajcik1974Phys.Rev.D1777,Forest1995Phys.Rev.C568}.
Within the first framework, the experimental triton binding energy can be nicely reproduced based on  covariant three-body spectator (or Gross) equations without explicitly adding $3N$ forces~\cite{Stadler1997Phys.Rev.Lett.26}.
For larger systems, the recent relativistic Brueckner Hartree-Fock (RBHF) calculations provide a very satisfactory description of nuclei up to $A\sim$ 40 without the inclusion of $3N$ interactions~\cite{Shen2019Prog.Part.Nucl.Phys.103713}.
Within the second framework, the concept of interparticle potentials, which is extremely useful in nonrelativistic many-body calculations, has been extended to a covariant fashion through the Poincar\'e group theory~\cite{Bakamjian1953Phys.Rev.1300, Krajcik1974Phys.Rev.D1777,Forest1995Phys.Rev.C568}.
The resultant relativistic Hamiltonian consists of relativistic kinetic energies, two- and three-body interactions as well as their Lorentz-boost corrections.
Using phenomenologically parametrized $NN$ and $3N$ interactions, variational Monte Carlo (VMC) calculations showed that the two-body boost interactions give repulsive contributions to the $^3$H and $^4$He ground states~\cite{Carlson1993Phys.Rev.C484,Forest1995Phys.Rev.C576,Forest1999Phys.Rev.C014002}.
This appears to be opposite to the results based on field theories, where the relativistic effects provide attractive contributions~\cite{Rupp1992Phys.Rev.C2133,Stadler1997Phys.Rev.Lett.26,Shen2019Prog.Part.Nucl.Phys.103713}.

Apparently, we have not yet arrived at a satisfactory understanding of the relativistic effects and their interplay with the $3N$ interactions in nuclear many-body systems.
Generally speaking, both relativity and three-body interactions influence the nuclear binding by inducing nonlocalities in the nuclear interactions.
Of course, the two effects are often intermingled, so  controversies could be caused by either different frameworks employed to consider the relativistic effects or disparate phenomenological models for the $3N$ interactions.
Therefore, it requires a framework that treats the relativistic effects as well as the $NN$ and $3N$ interactions in a unified and consistent way.

Since the advent of the effective field theory (EFT) paradigm in the early 1990s~\cite{Weinberg1990Phys.Lett.B288,Weinberg1991Nucl.Phys.B3}, a variety of studies have been carried out for the construction of $NN$ and $3N$ potentials within the nuclear EFT approach~\cite{Ordonez1996Phys.Rev.C2086,Entem2003Phys.Rev.C41001,Gezerlis2013Phys.Rev.Lett.32501,Epelbaum2015Phys.Rev.Lett.122301,Piarulli2018Phys.Rev.Lett.52503,Schiavilla2021Phys.Rev.C054003} (for reviews see Refs.~\cite{Bedaque2002Ann.Rev.Nucl.Part.Sci.339,Epelbaum2009Rev.Mod.Phys.1773,Machleidt2011Phys.Rep.1}).
To properly set up an EFT, one introduces a heavy scale  to separate the domain of applicability and the unresolved high-energy processes, whose effects are subsumed in the values of low-energy constants (LECs).
For nuclei in which the typical momenta of nucleons are much smaller than the pion mass, such a scale can be taken to be smaller than the pion mass.
This brings the pionless EFT~\cite{Chen1999Nucl.Phys.A386,Hammer2020Rev.Mod.Phys.025004}, where the low-energy properties of nuclei are not sensitive to the details associated with pion or other meson exchanges and, thus, can be described by an effective Lagrangian consisting of  only contact interactions between two or more nucleons.

The pionless EFT provides consistent $NN$ and $3N$ interactions, which emerge on an equal footing under a systematic expansion controlled by the power counting.
At leading order (LO), the power counting and renormalization-group invariance of pionless EFT are well understood up to few-body level~\cite{Yang2020Eur.Phys.J.A96}.
The LO pionless EFT has achieved great success in the description of $A\leq 4$ nuclei~\cite{Kirscher2015Phys.Rev.C054002,Contessi2017Phys.Lett.B839} and the explanation of the correlations among few-body observables, such as the Phillips~\cite{Phillips1968Nucl.Phys.A209216} and Tjon~\cite{Tjon1975Phys.Lett.B217220} lines.
Recently, VMC calculations based on artificial neural networks (ANNs) are carried out to solve the nuclear Hamiltonian derived from the LO pionless EFT~\cite{Adams2021Phys.Rev.Lett.022502}, giving accurate ground-state energies comparable to the virtually exact Green's function Monte Carlo (GFMC) results.
Therefore, the pionless EFT provides an ideal framework  to study the interplay between relativistic effects and $3N$ interactions for at least light nuclei.

In this work, a relativistic nuclear Hamiltonian is derived based on the leading-order covariant pionless EFT containing consistent relativistic and $3N$ potentials, and the ground-state properties of $A\le 4$ nuclei are computed with the VMC approach by introducing a symmetry-based ANN representation of the wave function and efficient stochastic sampling and optimization algorithms.
The relativistic effects and $3N$ interactions are analysed consistently with the calculated ground-state energies of $A\le 4$ nuclei, providing the first unified and consistent study of the interplay between relativistic effects and $3N$ interactions.


{\it Relativistic Hamiltonian.}---Within LO pionless EFT, we start from a manifestly covariant four-fermion contact Lagrangian,
\begin{eqnarray}
  \mathcal{L}^{(0)}_{NN}&=&-\frac{1}{2}\left[{C}_S(\overline{\psi}\psi)(\overline{\psi}\psi)+{C}_V(\overline{\psi}\gamma_\mu \psi)(\overline{\psi}\gamma^\mu \psi)\right.\nonumber\\
  &+&{C}_{P}(\overline{\psi}\gamma_5\psi)(\overline{\psi}\gamma_5\psi)+{C}_{AV}(\overline{\psi}\gamma_5\gamma_\mu \psi)(\overline{\psi}\gamma_5\gamma^\mu \psi)\nonumber\\
  &+&\left.{C}_T(\overline{\psi}\sigma_{\mu\nu}\psi)(\overline{\psi}\sigma^{\mu\nu}\psi)\right], \label{Lag}
\end{eqnarray}
with the nucleon field $\psi$ and LECs ${C}_S$, ${C}_V$, ${C}_{P}$, ${C}_{AV}$, and ${C}_{T}$.
In principle, the effects associated with virtual nucleon-antinucleon pairs can be absorbed in the LECs, since the creation of such virtual heavy pairs is a short-range process~\cite{Hammer2020Rev.Mod.Phys.025004}.
As a result, the nucleon field can be written in momentum space in terms of the free Dirac spinors with positive energies $u(\bm p, s)$, which read, by expanding in powers of $\bm p/m_N$,
\begin{equation}
  u(\bm p, s)=u_0(s)+u_1(\frac{\bm p}{m_N},s)+u_2(\frac{\bm p^2}{m_N^2}, s)+\cdots
\end{equation}
Only the leading term $u_0(s)$ should be considered at LO to achieve a renormalizable formulation for two-nucleon sectors, while other terms are identified as higher-order corrections~\cite{Epelbaum2012Phys.Lett.B338}.
Inserting the expansion of the nucleon field to the Lagrangian (\ref{Lag}), the LO $NN$ interaction can be easily obtained,
\begin{equation}
	\label{Eq.V2N}
  {V}_{NN}(\bm p_i', \bm p_j',\bm p_i, \bm p_j)=\mathcal{N}_{p_i'}\mathcal{N}_{p_i}\mathcal{N}_{p_j'}\mathcal{N}_{p_j}
  \left[{C}_1+{C}_2\bm\sigma_i\cdot\bm\sigma_j\right]
\end{equation}
with $\mathcal{N}_p=(1+ \bm p^2/m^2_N)^{-1/4}$. The LECs are reduced into two independent ones, i.e., ${C}_1={C}_S+{C}_V$ and ${C}_2=-{C}_{AV}+2{C}_T$, which corresponds to the central and spin-spin interactions, respectively.
This interaction is Lorentz invariant, and is consistent with a three-dimensional reduction of the Bethe-Salpeter equation~\cite{[See Supplemental Material for the theoretical derivation of the relativistic Hamiltonian as well as the performance of the symmetry-based artificial neural-network ansatz and the Gradient-Adaptive algorithm]supp}.

The ultraviolet divergences could be controlled by regularization and renormalization with a Gaussian regulator $\exp(q_{ij}^2/\Lambda^2)$ suppressing the momentum transfers $q_{ij}=\frac{1}{2}(p_i'-p_j')-\frac{1}{2}(p_i-p_j)$ above the cutoff $\Lambda$.
The relativistic effects are induced by the nonlocal parts of the interactions with the $\mathcal{N}_{p}$ factors and the timelike component $q^0_{ij}$ in the regulator.
We consider the dominant relativistic effects by keeping the leading terms in the expansions in powers of $\bm{p}/m_N$.
As a result, the regularized LO $2N$ interactions read~\cite{supp}
\begin{equation}\label{Eq.V2N.LO}
  {V}^\Lambda_{NN}=({C}_1+{C}_2\bm\sigma_i\cdot\bm\sigma_j)\mathrm{e}^{-\bm q_{ij}^2/\Lambda^2}\cdot\left(1+ V_\mathrm{b} + V_\mathrm{t} \right),
\end{equation}
where the relativistic corrections are clearly seen from the two terms $V_\mathrm{b}$ and $V_\mathrm{t}$,
\begin{equation}
  V_\mathrm{b} = -\frac{\bm P_{ij}^2}{4m_N^2} + \frac{(\bm P_{ij}\cdot\bm q_{ij})^2}{4m_N^2\Lambda^2},\quad V_\mathrm{t} = -\frac{\bm q_{ij}^2+\bm k_{ij}^2}{4m_N^2}.
\end{equation}
The first term $V_\mathrm{b}$ depends upon the total momentum of the two nucleons $\bm P_{ij}=\bm p_i+\bm p_j=\bm p_i'+\bm p_j'$ and provides the so-called boost interaction, which is consistent with the previous derivation from relativistic Hamiltonian dynamics~\cite{Krajcik1974Phys.Rev.D1777,Carlson1993Phys.Rev.C484}.
In contrast, the second term $V_\mathrm{t}$ depends on the momentum transfers in the direct $\bm q_{ij}=\bm p_i'-\bm p_i$ and exchange $\bm k_{ij}=\bm p_i'-\bm p_j$ channels,
and, thus, is called ``transfer'' interaction hereafter.

With Fourier transformation, one can finally express the relativistic Hamiltonian in coordinate space~\cite{supp},
\begin{eqnarray}\label{Eq.Rel.Ham.Coor}
\hat{H}_\mathrm{LO} & = &\sum_{i=1}^A\left[(m_N^2-\nabla_i^2)^{1/2}-m_N\right] \nonumber \\
&+  & \sum_{i<j}^A(C_1+C_{2}\bm\sigma_i\cdot\bm\sigma_j) \left( 1 + V_\mathrm{b}+ V_\mathrm{t}\right) \mathrm{e}^{-\frac{\Lambda^2}{4}\bm r_{ij}^2}.
\end{eqnarray}
Here, the relativistic kinetic energy is considered, and the LECs have been redefined with a factor from the Fourier transformation.
The boost and transfer interactions in coordinate space read,
\begin{equation} \label{Eq.Vboost}
V_\mathrm{b}(\bm r_{ij})=
	-\frac{\hat{\bm P}_{ij}^2}{8m_N^2}-\frac{\Lambda^2}{16m_N^2}(\hat{\bm P}_{ij}\cdot\bm r_{ij})^2,
\end{equation}
\begin{equation} \label{Eq.Vtrans}
V_\mathrm{t}(\bm r_{ij})=-\frac{\Lambda^2}{4m_N^2}
	\left[\left(3-\frac{\Lambda^2}{2}\bm r_{ij}^2\right)+2\mathrm{i}\bm r_{ij}\cdot\hat{\bm p}_{ij}
	+4\frac{\hat{\bm p}_{ij}^2}{\Lambda^2}\right],
\end{equation}
where the total momentum operator of the $i$th and $j$th nucleons is $\hat{\bm P}_{ij}=-\mathrm{i}(\bm\nabla_i+\bm\nabla_j)$ and the relative momentum operator is $\hat{\bm p}_{ij}=-\frac{\mathrm{i}}{2}(\bm\nabla_i-\bm\nabla_j)$.

{\it Variational Monte Carlo.}---The relativistic Hamiltonian (\ref{Eq.Rel.Ham.Coor}) is solved with the VMC approach, where one optimizes a parametrized variational wave function $\Psi_V$ by minimizing the total energy~\cite{Carlson1991}
\begin{equation}
  \frac{\langle\Psi_V|H|\Psi_V\rangle}{\langle\Psi_V|\Psi_V\rangle}=E_V.
\end{equation}
The brackets here contain a sum and an integral over the spin-isospin and the $3A$ spatial coordinates, respectively, and the integral can be evaluated with the Metropolis Monte Carlo algorithm~\cite{Metropolis1953J.Chem.Phys.1087}.

In the present work, the variational wave function is expressed in the following ANN representation
\begin{equation}\label{Eq.ANN}
  |\left. \Psi_V^{\rm ANN}\right\rangle=\exp\left[\mathcal{U}(\overline{r}_1, \ldots,\overline{r}_A; r_{12},\ldots,r_{A-1,A})\right]|\Phi\rangle,
\end{equation}
where $|\Phi\rangle$ is a suitable mean-field state, and $\mathcal{U}$ is a real-valued correlating factor parametrized by an ANN.
For the $s$-shell nuclei considered in this work, the mean-field part is taken to be $|\Phi_\mathrm{^3H}\rangle=\mathcal{A}|\uparrow_n\downarrow_n\uparrow_p\rangle$ and $|\Phi_\mathrm{^4He}\rangle=\mathcal{A}|\uparrow_n\downarrow_n\uparrow_p\downarrow_p\rangle$, with $\mathcal{A}$ being the antisymmetrization operator~\cite{LomnitzAdler1981Nucl.Phys.A399}.
Different from the previous work~\cite{Adams2021Phys.Rev.Lett.022502}, the correlating factor $\mathcal{U}$ here depends only on the distances from the nucleons to the center of mass of the nucleus $\overline{r}_{i}=|\bm r_i-\bm r_\mathrm{cm}|$ and the distances between two nucleons $r_{ij}=|\bm r_i-\bm r_j|$.
This choice is nicely compatible with the translational and rotational invariance of the Hamiltonian~(\ref{Eq.Rel.Ham.Coor}), which leads to a spatially rotation-invariant ground state.
Consequently, the present symmetry-based ANN can provide lower ground-state energies even by a smaller-scale ANN~\cite{supp}, which features about one order of magnitude fewer variational parameters than the network adopted in Ref.~\cite{Adams2021Phys.Rev.Lett.022502}.
The spin-dependent correlations are neglected in the present work, since we find that their contributions to the ground-state energies of $^{3}$H and $^{4}$He are quite small; similar to the nonrelativistic calculations~\cite{Adams2021Phys.Rev.Lett.022502}.

The ANN is comprised of four fully connected layers, including an input layer, two hidden layers, and an output layer.
The input layer has $A(A+1)/2$ nodes corresponding to the scalar variables of $\mathcal{U}$.
The output layer has one node giving the value of the correlating factor, where a Gaussian function is added to confine the nucleons within a finite volume, i.e., $\mathcal{U}\rightarrow\mathcal{U}-\alpha\sum_{i=1}^A\overline{r}_i^2$ with $\alpha=0.05$ as in Ref.~\cite{Adams2021Phys.Rev.Lett.022502}.
For the hidden layers, the differentiable softplus activation function~\cite{C.Dugas2001472478} is adopted, so the ground-state wave functions are differentiable with respect to the spatial coordinates.
The weights and biases connecting the layers serve as variational parameters, which are optimized by the stochastic reconfiguration (SR) algorithm~\cite{Sorella1998Phys.Rev.Lett.45584561,Sorella2005Phys.Rev.B241103}, but the learning rate $\gamma$ in the SR algorithm has to be carefully customized to achieve a stable and efficient convergence.
The AdaptiveEta algorithm proposed for nonrelativistic calculations~\cite{Adams2021Phys.Rev.Lett.022502} is not used here, because it relies on heuristic tests on the parameter change and could be very time-consuming, in particular for relativistic calculations due to the repetitive evaluations of the relativistic kinetic energy.
Therefore, we introduce the Gradient-Adaptive algorithm, in which the parameter change between two iterations are optimally controlled via a tradeoff between the stability and efficiency~\cite{supp}.

{\it Results and discussion.}---The LECs $C_1$ and $C_2$ in the relativistic Hamiltonian (\ref{Eq.Rel.Ham.Coor}) are determined by reproducing the experimental singlet $np$ scattering length and the $^2$H ground-state energy.
Their values at different cutoffs $\Lambda$ are listed in Table~\ref{Tab.LECs} in comparison with the nonrelativistic ones~\cite{Kirscher2015Phys.Rev.C054002}.
The $NN$ interactions are dominated by their central components, i.e., $C_1\gg C_2$.
In comparison with the nonrelativistic $C_1$ values, the relativistic ones are smaller in amplitude without the transfer interaction, while larger in amplitude with the transfer interaction.
As the relativistic kinetic energy is generally smaller than the nonrelativistic one, a fit without the transfer interaction to the same $^2$H ground-state energy leads to a less attractive $NN$ interaction.
However, the same fit with the transfer interaction leads to a more attractive $NN$ coupling strength as the transfer interaction is repulsive at short range [see Eq.~(\ref{Eq.Vtrans})] and, thus, tends to draw nucleons further apart.

\begin{table}[!htbp]
  \centering
  \caption{\label{Tab.LECs}Low-energy constants $C_1$ and $C_2$ (in GeV) at different cutoffs $\Lambda$ for the nonrelativistic LO Hamiltonian~\cite{Kirscher2015Phys.Rev.C054002} and the relativistic ones without and with ``transfer'' interactions.}
  \begin{ruledtabular}
  \begin{tabular}{ccccccc}
    $\Lambda$  & \multicolumn{2}{c}{Non-rel.~\cite{Kirscher2015Phys.Rev.C054002}}   & \multicolumn{2}{c}{Rel. w/o trans.} & \multicolumn{2}{c}{Rel.}\\
    \cline{2-3}\cline{4$\ $-5}\cline{6$\ $-7}
               & $C_{1}$  & $C_{2}$ & $C_1$  & $C_{2}$ & $C_1$  & $C_{2}$  \\ \hline
    2 fm$^{-1}$& -0.133   & -0.009  & -0.131 & -0.009  & -0.141 & -0.010   \\
    4 fm$^{-1}$& -0.488   & -0.017  & -0.462 & -0.016  & -0.595 & -0.027   \\
    6 fm$^{-1}$& -1.06    & -0.026  & -0.958 & -0.021  & -1.53  & -0.055   \\
    8 fm$^{-1}$& -1.87    & -0.035  & -1.58  & -0.024  & -2.88  & -0.086   \\
  \end{tabular}
  \end{ruledtabular}
\end{table}

\begin{figure}[!htbp]
  \centering
  \includegraphics[width=0.45\textwidth]{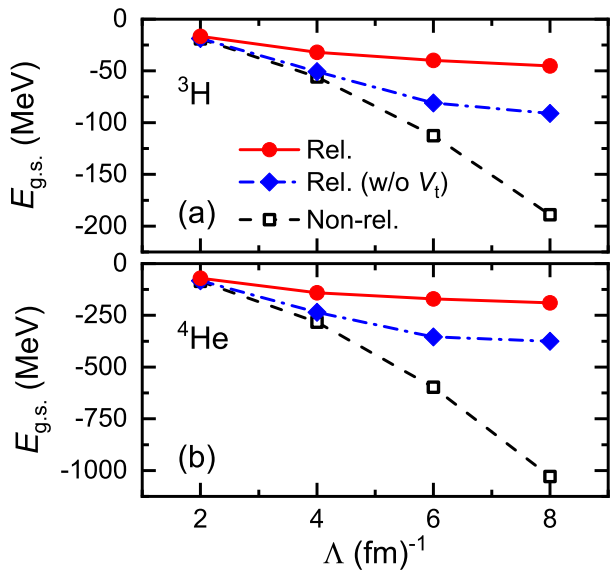}
  \caption{\label{Fig.no3NF} (Color online). The ground-state energies of $^3$H (a) and $^4$He (b) obtained with the nonrelativistic LO Hamiltonian and the relativistic ones without and with ``transfer'' interactions, as functions of cutoff $\Lambda$.}
\end{figure}

The determined $NN$ interactions can be used to calculate $A\geq3$ nuclei.
In Fig.~\ref{Fig.no3NF}, the ground-state energies of $^3$H and $^4$He are depicted at different cutoffs.
In the absence of $3N$ interactions, the ground-state energies of $^3$H and $^4$He given by the nonrelativistic Hamiltonian collapse as the cutoff $\Lambda\rightarrow\infty$, showing a lack of the renormalizability.
This nontrivial renormalization problem is well known as the ``Thomas collapse"~\cite{Thomas1935Phys.Rev.903} in LO pionless EFT, which, in the standard nonrelativistic formulations, is avoided by promoting a repulsive $3N$ interaction to LO~\cite{Bedaque1999Phys.Rev.Lett.463}.
In contrast, the relativistic results converge as the cutoff increases, so the ``Thomas collapse'' problem can be overcome by taking into account relativistic effects instead of introducing $3N$ interactions.

The boost interaction (\ref{Eq.Vboost}) indeed plays a similar role as a repulsive $3N$ interaction.
Taking $^3$H as an example, the term $\hat{\bm P}_{12}=\hat{\bm p}_1+\hat{\bm p}_2$ can be readily replaced with $-\hat{\bm p}_3$ using the center-of-mass condition, so the boost interaction (\ref{Eq.Vboost})  can be rewritten as a $3N$ form,
\begin{equation}
	V_\mathrm{b}(\hat{\bm p}_{3},\bm r_{12})=-\frac{\hat{\bm p}_{3}^2}{4m_N^2}-\frac{\Lambda^2}{16m_N^2}(\hat{\bm p}_{3}\cdot\bm r_{12})^2.
\end{equation}
For the transfer interaction $V_\mathrm{t}$ (\ref{Eq.Vtrans}), it has a short-ranged repulsive core, whose strength grows as the cutoff $\Lambda$ increases.
This would hinder the nucleons coming infinitely close to each other when $\Lambda\rightarrow\infty$ and, thus, avoids the Thomas collapse.

The relativistic effects would be even more interesting for the next-to-leading order (NLO), where a four-body interaction is surprisingly needed to prevent a collapse of the $^4$He ground state in nonrelativistic calculations~\cite{Bazak2019Phys.Rev.Lett.143001}.
Since the relativistic effects here rescue the renormalizability at leading order without three-body forces, it will be interesting to see similar effects at the NLO without four-body interactions.

Note that the calculated $^3$H and $^4$He ground states with the relativistic LO Hamiltonian are still too overbound, though not collapse, in comparison with the experiment.
In this sense, a repulsive $3N$ interaction may be needed to alleviate the overbinding, but the underlying physics is different from the nonrelativistic case, where the $3N$ interaction is required by the renormalizability.

\begin{figure}[!htbp]
  \centering
  \includegraphics[width=0.45\textwidth]{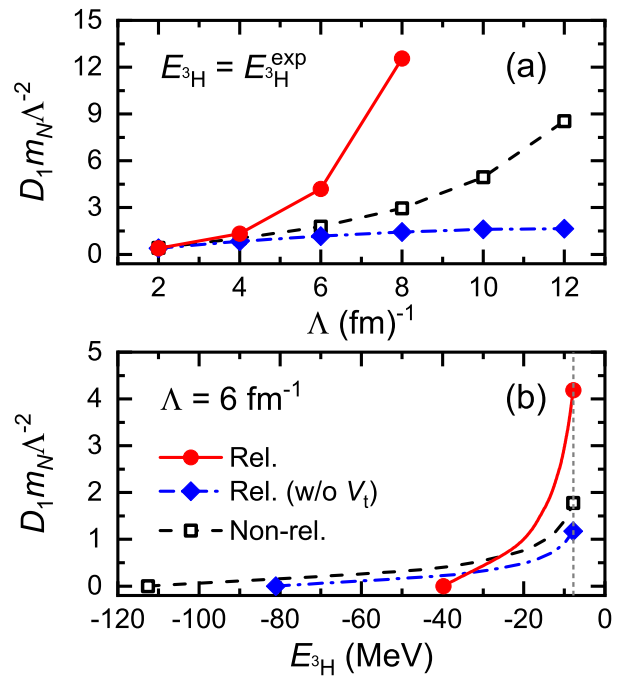}
  \caption{\label{Fig.3NF} (Color online). The dimensionless strengths $D_1m_N\Lambda^{-2}$ of the three-nucleon interactions in the nonrelativistic
  LO Hamiltonian and the relativistic ones without and with ``transfer'' interactions: (a) as functions of the cutoff $\Lambda$ by fixing the $^3$H ground-state energy at the experimental value; (b) as functions of the $^3$H ground-state energy by fixing the cutoff at $\Lambda = 6 $ fm$^{-1}$. The vertical dashed line in panel (b) denotes the experimental ground-state energy of $^3$H.}
\end{figure}

In analogy with the covariant $NN$ interactions, the regularized LO $3N$ interactions can also be derived~\cite{supp}, but the corresponding relativistic corrections are expected to be small as compared to the $NN$ counterparts~\cite{Carlson1993Phys.Rev.C484}, so they are neglected in the following calculations.
In Fig.~\ref{Fig.3NF}(a), the dimensionless strengths $D_1m_N\Lambda^{-2}$ of the $3N$ interactions determined by the experimental $^3$H ground-state energy are depicted as functions of the cutoff $\Lambda$.
For the nonrelativistic Hamiltonian, the $3N$ interaction strength grows monotonically with the cutoff $\Lambda$, and this is a direct consequence of the Thomas Collapse.
For the relativistic Hamiltonian with only boost interactions, in contrast, the $3N$ interaction strength converges as the cutoff $\Lambda$ increases, and this should be associated with the convergent behavior of the $^3$H ground-state energies as $\Lambda\rightarrow\infty$ [see Fig.~\ref{Fig.no3NF}(a)].
By including the transfer interactions further, one may expect a further reduction of the $3N$ interaction strength since, as seen in Fig.~\ref{Fig.no3NF}(a), the transfer interactions correct the energy remarkably.
However, surprisingly, a strong enhancement of the strength is obtained.

To understand this behavior, the running of the $3N$ interaction strength with respect to the $^3$H ground-state energy is depicted in Fig.~\ref{Fig.3NF}(b) by taking $\Lambda=6$ fm$^{-1}$ as an example.
As the $^3$H ground-state energy approaching to the experimental value, the required strengths of the $3N$ interactions gradually increase in all cases, but the growing rate is significantly large when the relativistic transfer interactions are considered.
This means that a large $3N$ interaction strength does not necessarily correspond to a large contribution to the calculated energy.
Following the Hellmann-Feynman theorem~\cite{Hellmann1937,Feynman1939Phys.Rev.340},
\begin{equation}
  \frac{\mathrm{d}E_{^3\mathrm{H}}}{\mathrm{d}D_1}=\left\langle\Psi\left|\frac{\partial\hat{H}_\mathrm{LO}}{\partial D_1}\right|\Psi\right\rangle
  =\left\langle\Psi\left|\sum_{i<j<k}\hat{f}_{3N}(i,j,k)\right|\Psi\right\rangle,
\end{equation}
where  $\hat{f}_{3N}(i,j,k)=\sum_\mathrm{cyc}\mathrm{e}^{-\Lambda^2(r_{ij}^2+r_{ik}^2)/4}$ is the functional form of the $3N$ interaction between three nucleons $i,j,k$~\cite{supp}, one can see that the $3N$ interaction strength $D_1$ should generally be related to the average distances between nucleons, i.e., the further the nucleons are spread, the lower the probability of the three nucleons coming close to each other, and consequently, the stronger the required short-ranged $3N$ interaction to reproduce the experimental energy.
In the present relativistic LO Hamiltonian, the transfer interactions $V_\mathrm{t}$ tend to spread the nucleons, so it requires a stronger $3N$ interaction in order to reproduce the experimental energy.
This demonstrates a strong interplay between the relativistic effects and the $3N$ interactions in LO pionless EFT.
How do other high-order relativistic effects and/or pion degrees of freedom influence such an interplay would be an interesting question for future study.


In summary, we have derived a relativistic nuclear Hamiltonian based on the leading-order covariant pionless EFT containing consistent relativistic and $3N$ potentials, and developed the corresponding relativistic VMC approach using a symmetry-based ANN representation of the nuclear many-body wave function.
By determining the two-body LECs in the Hamiltonian with the experimental singlet $np$ scattering length and $^2$H binding energy,
the ground-state energies of $A\le 4$ nuclei are calculated with efficient stochastic sampling and optimization methods.
It is found that the problem of Thomas collapse at large cutoffs existing in nonrelativistic calculations can be avoided by taking into account relativistic effects instead of promoting a repulsive $3N$ interaction to leading order.
While the relativistic effects rescue the renormalizability at leading order, a repulsive $3N$ interaction may be needed to alleviate the overbinding of the calculated ground-state energies for $^3$H and $^4$He as compared to data.
Due to the strong repulsion of the two-body relativistic effects, the probability of finding adjacent nucleon triplets becomes lower.
This interplay hinders the impacts of $3N$ interactions on the ground-state energy, and may cause a large $3N$ interaction strength to reproduce the experimental  data at leading order.
The presented results open the avenue for a unified and consistent study on relativistic effects and three-body interactions in atomic nuclei.

\begin{acknowledgments}
This work has been supported in part by the National Key R\&D Program of China (Contracts No. 2017YFE0116700 and No. 2018YFA0404400), the National Natural Science Foundation of China (Grants No. 12070131001, No. 11875075, No. 11935003, No. 11975031, No. 12105004, and No. 12141501), and the High-performance Computing Platform of Peking University.
\end{acknowledgments}


\begin{thebibliography}{48}%
\makeatletter
\providecommand \@ifxundefined [1]{%
 \@ifx{#1\undefined}
}%
\providecommand \@ifnum [1]{%
 \ifnum #1\expandafter \@firstoftwo
 \else \expandafter \@secondoftwo
 \fi
}%
\providecommand \@ifx [1]{%
 \ifx #1\expandafter \@firstoftwo
 \else \expandafter \@secondoftwo
 \fi
}%
\providecommand \natexlab [1]{#1}%
\providecommand \enquote  [1]{``#1''}%
\providecommand \bibnamefont  [1]{#1}%
\providecommand \bibfnamefont [1]{#1}%
\providecommand \citenamefont [1]{#1}%
\providecommand \href@noop [0]{\@secondoftwo}%
\providecommand \href [0]{\begingroup \@sanitize@url \@href}%
\providecommand \@href[1]{\@@startlink{#1}\@@href}%
\providecommand \@@href[1]{\endgroup#1\@@endlink}%
\providecommand \@sanitize@url [0]{\catcode `\\12\catcode `\$12\catcode
  `\&12\catcode `\#12\catcode `\^12\catcode `\_12\catcode `\%12\relax}%
\providecommand \@@startlink[1]{}%
\providecommand \@@endlink[0]{}%
\providecommand \url  [0]{\begingroup\@sanitize@url \@url }%
\providecommand \@url [1]{\endgroup\@href {#1}{\urlprefix }}%
\providecommand \urlprefix  [0]{URL }%
\providecommand \Eprint [0]{\href }%
\providecommand \doibase [0]{http://dx.doi.org/}%
\providecommand \selectlanguage [0]{\@gobble}%
\providecommand \bibinfo  [0]{\@secondoftwo}%
\providecommand \bibfield  [0]{\@secondoftwo}%
\providecommand \translation [1]{[#1]}%
\providecommand \BibitemOpen [0]{}%
\providecommand \bibitemStop [0]{}%
\providecommand \bibitemNoStop [0]{.\EOS\space}%
\providecommand \EOS [0]{\spacefactor3000\relax}%
\providecommand \BibitemShut  [1]{\csname bibitem#1\endcsname}%
\let\auto@bib@innerbib\@empty
\bibitem [{\citenamefont {Friar}\ \emph {et~al.}(1984)\citenamefont {Friar},
  \citenamefont {Gibson},\ and\ \citenamefont
  {Payne}}]{Friar1984Annu.Rev.Nucl.Part.Sci.403}%
  \BibitemOpen
  \bibfield  {author} {\bibinfo {author} {\bibfnamefont {J.~L.}\ \bibnamefont
  {Friar}}, \bibinfo {author} {\bibfnamefont {B.~F.}\ \bibnamefont {Gibson}}, \
  and\ \bibinfo {author} {\bibfnamefont {G.~L.}\ \bibnamefont {Payne}},\ }\href
  {\doibase 10.1146/annurev.ns.34.120184.002155} {\bibfield  {journal}
  {\bibinfo  {journal} {Annu. Rev. Nucl. Part. Sci.}\ }\textbf {\bibinfo
  {volume} {34}},\ \bibinfo {pages} {403} (\bibinfo {year} {1984})}\BibitemShut
  {NoStop}%
\bibitem [{\citenamefont {Carlson}\ \emph {et~al.}(2015)\citenamefont
  {Carlson}, \citenamefont {Gandolfi}, \citenamefont {Pederiva}, \citenamefont
  {Pieper}, \citenamefont {Schiavilla}, \citenamefont {Schmidt},\ and\
  \citenamefont {Wiringa}}]{Carlson2015Rev.Mod.Phys.1067}%
  \BibitemOpen
  \bibfield  {author} {\bibinfo {author} {\bibfnamefont {J.}~\bibnamefont
  {Carlson}}, \bibinfo {author} {\bibfnamefont {S.}~\bibnamefont {Gandolfi}},
  \bibinfo {author} {\bibfnamefont {F.}~\bibnamefont {Pederiva}}, \bibinfo
  {author} {\bibfnamefont {S.~C.}\ \bibnamefont {Pieper}}, \bibinfo {author}
  {\bibfnamefont {R.}~\bibnamefont {Schiavilla}}, \bibinfo {author}
  {\bibfnamefont {K.~E.}\ \bibnamefont {Schmidt}}, \ and\ \bibinfo {author}
  {\bibfnamefont {R.~B.}\ \bibnamefont {Wiringa}},\ }\href {\doibase
  10.1103/RevModPhys.87.1067} {\bibfield  {journal} {\bibinfo  {journal} {Rev.
  Mod. Phys.}\ }\textbf {\bibinfo {volume} {87}},\ \bibinfo {pages} {1067}
  (\bibinfo {year} {2015})}\BibitemShut {NoStop}%
\bibitem [{\citenamefont {Lonardoni}\ \emph {et~al.}(2018)\citenamefont
  {Lonardoni}, \citenamefont {Carlson}, \citenamefont {Gandolfi}, \citenamefont
  {Lynn}, \citenamefont {Schmidt}, \citenamefont {Schwenk},\ and\ \citenamefont
  {Wang}}]{Lonardoni2018Phys.Rev.Lett.122502}%
  \BibitemOpen
  \bibfield  {author} {\bibinfo {author} {\bibfnamefont {D.}~\bibnamefont
  {Lonardoni}}, \bibinfo {author} {\bibfnamefont {J.}~\bibnamefont {Carlson}},
  \bibinfo {author} {\bibfnamefont {S.}~\bibnamefont {Gandolfi}}, \bibinfo
  {author} {\bibfnamefont {J.~E.}\ \bibnamefont {Lynn}}, \bibinfo {author}
  {\bibfnamefont {K.~E.}\ \bibnamefont {Schmidt}}, \bibinfo {author}
  {\bibfnamefont {A.}~\bibnamefont {Schwenk}}, \ and\ \bibinfo {author}
  {\bibfnamefont {X.~B.}\ \bibnamefont {Wang}},\ }\href {\doibase
  10.1103/PhysRevLett.120.122502} {\bibfield  {journal} {\bibinfo  {journal}
  {Phys. Rev. Lett.}\ }\textbf {\bibinfo {volume} {120}},\ \bibinfo {pages}
  {122502} (\bibinfo {year} {2018})}\BibitemShut {NoStop}%
\bibitem [{\citenamefont {Ren}\ \emph {et~al.}(2018)\citenamefont {Ren},
  \citenamefont {Li}, \citenamefont {Geng}, \citenamefont {Long}, \citenamefont
  {Ring},\ and\ \citenamefont {Meng}}]{Ren2018Chin.Phys.C14103}%
  \BibitemOpen
  \bibfield  {author} {\bibinfo {author} {\bibfnamefont {X.-L.}\ \bibnamefont
  {Ren}}, \bibinfo {author} {\bibfnamefont {K.-W.}\ \bibnamefont {Li}},
  \bibinfo {author} {\bibfnamefont {L.-S.}\ \bibnamefont {Geng}}, \bibinfo
  {author} {\bibfnamefont {B.}~\bibnamefont {Long}}, \bibinfo {author}
  {\bibfnamefont {P.}~\bibnamefont {Ring}}, \ and\ \bibinfo {author}
  {\bibfnamefont {J.}~\bibnamefont {Meng}},\ }\href
  {http://stacks.iop.org/1674-1137/42/i=1/a=014103} {\bibfield  {journal}
  {\bibinfo  {journal} {Chin. Phys. C}\ }\textbf {\bibinfo {volume} {42}},\
  \bibinfo {pages} {014103} (\bibinfo {year} {2018})}\BibitemShut {NoStop}%
\bibitem [{\citenamefont {Lu}\ \emph {et~al.}(2022)\citenamefont {Lu},
  \citenamefont {Wang}, \citenamefont {Xiao}, \citenamefont {Geng},
  \citenamefont {Meng},\ and\ \citenamefont
  {Ring}}]{Lu2022Phys.Rev.Lett.142002}%
  \BibitemOpen
  \bibfield  {author} {\bibinfo {author} {\bibfnamefont {J.-X.}\ \bibnamefont
  {Lu}}, \bibinfo {author} {\bibfnamefont {C.-X.}\ \bibnamefont {Wang}},
  \bibinfo {author} {\bibfnamefont {Y.}~\bibnamefont {Xiao}}, \bibinfo {author}
  {\bibfnamefont {L.-S.}\ \bibnamefont {Geng}}, \bibinfo {author}
  {\bibfnamefont {J.}~\bibnamefont {Meng}}, \ and\ \bibinfo {author}
  {\bibfnamefont {P.}~\bibnamefont {Ring}},\ }\href {\doibase
  10.1103/PhysRevLett.128.142002} {\bibfield  {journal} {\bibinfo  {journal}
  {Phys. Rev. Lett.}\ }\textbf {\bibinfo {volume} {128}},\ \bibinfo {pages}
  {142002} (\bibinfo {year} {2022})}\BibitemShut {NoStop}%
\bibitem [{\citenamefont {Brockmann}\ and\ \citenamefont
  {Machleidt}(1984)}]{Brockmann1984Phys.Lett.B283}%
  \BibitemOpen
  \bibfield  {author} {\bibinfo {author} {\bibfnamefont {R.}~\bibnamefont
  {Brockmann}}\ and\ \bibinfo {author} {\bibfnamefont {R.}~\bibnamefont
  {Machleidt}},\ }\href {\doibase https://doi.org/10.1016/0370-2693(84)90407-6}
  {\bibfield  {journal} {\bibinfo  {journal} {Phys. Lett. B}\ }\textbf
  {\bibinfo {volume} {149}},\ \bibinfo {pages} {283} (\bibinfo {year}
  {1984})}\BibitemShut {NoStop}%
\bibitem [{\citenamefont {Brockmann}\ and\ \citenamefont
  {Machleidt}(1990)}]{Brockmann1990Phys.Rev.C1965}%
  \BibitemOpen
  \bibfield  {author} {\bibinfo {author} {\bibfnamefont {R.}~\bibnamefont
  {Brockmann}}\ and\ \bibinfo {author} {\bibfnamefont {R.}~\bibnamefont
  {Machleidt}},\ }\href {\doibase 10.1103/PhysRevC.42.1965} {\bibfield
  {journal} {\bibinfo  {journal} {Phys. Rev. C}\ }\textbf {\bibinfo {volume}
  {42}},\ \bibinfo {pages} {1965} (\bibinfo {year} {1990})}\BibitemShut
  {NoStop}%
\bibitem [{\citenamefont {Stadler}\ and\ \citenamefont
  {Gross}(1997)}]{Stadler1997Phys.Rev.Lett.26}%
  \BibitemOpen
  \bibfield  {author} {\bibinfo {author} {\bibfnamefont {A.}~\bibnamefont
  {Stadler}}\ and\ \bibinfo {author} {\bibfnamefont {F.}~\bibnamefont
  {Gross}},\ }\href {\doibase 10.1103/PhysRevLett.78.26} {\bibfield  {journal}
  {\bibinfo  {journal} {Phys. Rev. Lett.}\ }\textbf {\bibinfo {volume} {78}},\
  \bibinfo {pages} {26} (\bibinfo {year} {1997})}\BibitemShut {NoStop}%
\bibitem [{\citenamefont {Forest}\ \emph {et~al.}(1999)\citenamefont {Forest},
  \citenamefont {Pandharipande},\ and\ \citenamefont
  {Arriaga}}]{Forest1999Phys.Rev.C014002}%
  \BibitemOpen
  \bibfield  {author} {\bibinfo {author} {\bibfnamefont {J.~L.}\ \bibnamefont
  {Forest}}, \bibinfo {author} {\bibfnamefont {V.~R.}\ \bibnamefont
  {Pandharipande}}, \ and\ \bibinfo {author} {\bibfnamefont {A.}~\bibnamefont
  {Arriaga}},\ }\href {\doibase 10.1103/PhysRevC.60.014002} {\bibfield
  {journal} {\bibinfo  {journal} {Phys. Rev. C}\ }\textbf {\bibinfo {volume}
  {60}},\ \bibinfo {pages} {014002} (\bibinfo {year} {1999})}\BibitemShut
  {NoStop}%
\bibitem [{\citenamefont {Shen}\ \emph {et~al.}(2019)\citenamefont {Shen},
  \citenamefont {Liang}, \citenamefont {Long}, \citenamefont {Meng},\ and\
  \citenamefont {Ring}}]{Shen2019Prog.Part.Nucl.Phys.103713}%
  \BibitemOpen
  \bibfield  {author} {\bibinfo {author} {\bibfnamefont {S.}~\bibnamefont
  {Shen}}, \bibinfo {author} {\bibfnamefont {H.}~\bibnamefont {Liang}},
  \bibinfo {author} {\bibfnamefont {W.~H.}\ \bibnamefont {Long}}, \bibinfo
  {author} {\bibfnamefont {J.}~\bibnamefont {Meng}}, \ and\ \bibinfo {author}
  {\bibfnamefont {P.}~\bibnamefont {Ring}},\ }\href {\doibase
  https://doi.org/10.1016/j.ppnp.2019.103713} {\bibfield  {journal} {\bibinfo
  {journal} {Prog. Part. Nucl. Phys.}\ }\textbf {\bibinfo {volume} {109}},\
  \bibinfo {pages} {103713} (\bibinfo {year} {2019})}\BibitemShut {NoStop}%
\bibitem [{\citenamefont {Rupp}\ and\ \citenamefont
  {Tjon}(1992)}]{Rupp1992Phys.Rev.C2133}%
  \BibitemOpen
  \bibfield  {author} {\bibinfo {author} {\bibfnamefont {G.}~\bibnamefont
  {Rupp}}\ and\ \bibinfo {author} {\bibfnamefont {J.~A.}\ \bibnamefont
  {Tjon}},\ }\href {\doibase 10.1103/PhysRevC.45.2133} {\bibfield  {journal}
  {\bibinfo  {journal} {Phys. Rev. C}\ }\textbf {\bibinfo {volume} {45}},\
  \bibinfo {pages} {2133} (\bibinfo {year} {1992})}\BibitemShut {NoStop}%
\bibitem [{\citenamefont {Bakamjian}\ and\ \citenamefont
  {Thomas}(1953)}]{Bakamjian1953Phys.Rev.1300}%
  \BibitemOpen
  \bibfield  {author} {\bibinfo {author} {\bibfnamefont {B.}~\bibnamefont
  {Bakamjian}}\ and\ \bibinfo {author} {\bibfnamefont {L.~H.}\ \bibnamefont
  {Thomas}},\ }\href {\doibase 10.1103/PhysRev.92.1300} {\bibfield  {journal}
  {\bibinfo  {journal} {Phys. Rev.}\ }\textbf {\bibinfo {volume} {92}},\
  \bibinfo {pages} {1300} (\bibinfo {year} {1953})}\BibitemShut {NoStop}%
\bibitem [{\citenamefont {Krajcik}\ and\ \citenamefont
  {Foldy}(1974)}]{Krajcik1974Phys.Rev.D1777}%
  \BibitemOpen
  \bibfield  {author} {\bibinfo {author} {\bibfnamefont {R.~A.}\ \bibnamefont
  {Krajcik}}\ and\ \bibinfo {author} {\bibfnamefont {L.~L.}\ \bibnamefont
  {Foldy}},\ }\href {\doibase 10.1103/PhysRevD.10.1777} {\bibfield  {journal}
  {\bibinfo  {journal} {Phys. Rev. D}\ }\textbf {\bibinfo {volume} {10}},\
  \bibinfo {pages} {1777} (\bibinfo {year} {1974})}\BibitemShut {NoStop}%
\bibitem [{\citenamefont {Forest}\ \emph
  {et~al.}(1995{\natexlab{a}})\citenamefont {Forest}, \citenamefont
  {Pandharipande},\ and\ \citenamefont {Friar}}]{Forest1995Phys.Rev.C568}%
  \BibitemOpen
  \bibfield  {author} {\bibinfo {author} {\bibfnamefont {J.~L.}\ \bibnamefont
  {Forest}}, \bibinfo {author} {\bibfnamefont {V.~R.}\ \bibnamefont
  {Pandharipande}}, \ and\ \bibinfo {author} {\bibfnamefont {J.~L.}\
  \bibnamefont {Friar}},\ }\href {\doibase 10.1103/PhysRevC.52.568} {\bibfield
  {journal} {\bibinfo  {journal} {Phys. Rev. C}\ }\textbf {\bibinfo {volume}
  {52}},\ \bibinfo {pages} {568} (\bibinfo {year}
  {1995}{\natexlab{a}})}\BibitemShut {NoStop}%
\bibitem [{\citenamefont {Carlson}\ \emph {et~al.}(1993)\citenamefont
  {Carlson}, \citenamefont {Pandharipande},\ and\ \citenamefont
  {Schiavilla}}]{Carlson1993Phys.Rev.C484}%
  \BibitemOpen
  \bibfield  {author} {\bibinfo {author} {\bibfnamefont {J.}~\bibnamefont
  {Carlson}}, \bibinfo {author} {\bibfnamefont {V.~R.}\ \bibnamefont
  {Pandharipande}}, \ and\ \bibinfo {author} {\bibfnamefont {R.}~\bibnamefont
  {Schiavilla}},\ }\href {\doibase 10.1103/PhysRevC.47.484} {\bibfield
  {journal} {\bibinfo  {journal} {Phys. Rev. C}\ }\textbf {\bibinfo {volume}
  {47}},\ \bibinfo {pages} {484} (\bibinfo {year} {1993})}\BibitemShut
  {NoStop}%
\bibitem [{\citenamefont {Forest}\ \emph
  {et~al.}(1995{\natexlab{b}})\citenamefont {Forest}, \citenamefont
  {Pandharipande}, \citenamefont {Carlson},\ and\ \citenamefont
  {Schiavilla}}]{Forest1995Phys.Rev.C576}%
  \BibitemOpen
  \bibfield  {author} {\bibinfo {author} {\bibfnamefont {J.~L.}\ \bibnamefont
  {Forest}}, \bibinfo {author} {\bibfnamefont {V.~R.}\ \bibnamefont
  {Pandharipande}}, \bibinfo {author} {\bibfnamefont {J.}~\bibnamefont
  {Carlson}}, \ and\ \bibinfo {author} {\bibfnamefont {R.}~\bibnamefont
  {Schiavilla}},\ }\href {\doibase 10.1103/PhysRevC.52.576} {\bibfield
  {journal} {\bibinfo  {journal} {Phys. Rev. C}\ }\textbf {\bibinfo {volume}
  {52}},\ \bibinfo {pages} {576} (\bibinfo {year}
  {1995}{\natexlab{b}})}\BibitemShut {NoStop}%
\bibitem [{\citenamefont {Weinberg}(1990)}]{Weinberg1990Phys.Lett.B288}%
  \BibitemOpen
  \bibfield  {author} {\bibinfo {author} {\bibfnamefont {S.}~\bibnamefont
  {Weinberg}},\ }\href {\doibase 10.1016/0370-2693(90)90938-3} {\bibfield
  {journal} {\bibinfo  {journal} {Phys. Lett. B}\ }\textbf {\bibinfo {volume}
  {251}},\ \bibinfo {pages} {288} (\bibinfo {year} {1990})}\BibitemShut
  {NoStop}%
\bibitem [{\citenamefont {Weinberg}(1991)}]{Weinberg1991Nucl.Phys.B3}%
  \BibitemOpen
  \bibfield  {author} {\bibinfo {author} {\bibfnamefont {S.}~\bibnamefont
  {Weinberg}},\ }\href {\doibase 10.1016/0550-3213(91)90231-L} {\bibfield
  {journal} {\bibinfo  {journal} {Nucl. Phys. B}\ }\textbf {\bibinfo {volume}
  {363}},\ \bibinfo {pages} {3} (\bibinfo {year} {1991})}\BibitemShut {NoStop}%
\bibitem [{\citenamefont {Ord\'o\~nez}\ \emph {et~al.}(1996)\citenamefont
  {Ord\'o\~nez}, \citenamefont {Ray},\ and\ \citenamefont {van
  Kolck}}]{Ordonez1996Phys.Rev.C2086}%
  \BibitemOpen
  \bibfield  {author} {\bibinfo {author} {\bibfnamefont {C.}~\bibnamefont
  {Ord\'o\~nez}}, \bibinfo {author} {\bibfnamefont {L.}~\bibnamefont {Ray}}, \
  and\ \bibinfo {author} {\bibfnamefont {U.}~\bibnamefont {van Kolck}},\ }\href
  {\doibase 10.1103/PhysRevC.53.2086} {\bibfield  {journal} {\bibinfo
  {journal} {Phys. Rev. C}\ }\textbf {\bibinfo {volume} {53}},\ \bibinfo
  {pages} {2086} (\bibinfo {year} {1996})}\BibitemShut {NoStop}%
\bibitem [{\citenamefont {Entem}\ and\ \citenamefont
  {Machleidt}(2003)}]{Entem2003Phys.Rev.C41001}%
  \BibitemOpen
  \bibfield  {author} {\bibinfo {author} {\bibfnamefont {D.~R.}\ \bibnamefont
  {Entem}}\ and\ \bibinfo {author} {\bibfnamefont {R.}~\bibnamefont
  {Machleidt}},\ }\href {\doibase 10.1103/PhysRevC.68.041001} {\bibfield
  {journal} {\bibinfo  {journal} {Phys. Rev. C}\ }\textbf {\bibinfo {volume}
  {68}},\ \bibinfo {pages} {041001} (\bibinfo {year} {2003})}\BibitemShut
  {NoStop}%
\bibitem [{\citenamefont {Gezerlis}\ \emph {et~al.}(2013)\citenamefont
  {Gezerlis}, \citenamefont {Tews}, \citenamefont {Epelbaum}, \citenamefont
  {Gandolfi}, \citenamefont {Hebeler}, \citenamefont {Nogga},\ and\
  \citenamefont {Schwenk}}]{Gezerlis2013Phys.Rev.Lett.32501}%
  \BibitemOpen
  \bibfield  {author} {\bibinfo {author} {\bibfnamefont {A.}~\bibnamefont
  {Gezerlis}}, \bibinfo {author} {\bibfnamefont {I.}~\bibnamefont {Tews}},
  \bibinfo {author} {\bibfnamefont {E.}~\bibnamefont {Epelbaum}}, \bibinfo
  {author} {\bibfnamefont {S.}~\bibnamefont {Gandolfi}}, \bibinfo {author}
  {\bibfnamefont {K.}~\bibnamefont {Hebeler}}, \bibinfo {author} {\bibfnamefont
  {A.}~\bibnamefont {Nogga}}, \ and\ \bibinfo {author} {\bibfnamefont
  {A.}~\bibnamefont {Schwenk}},\ }\href {\doibase
  10.1103/PhysRevLett.111.032501} {\bibfield  {journal} {\bibinfo  {journal}
  {Phys. Rev. Lett.}\ }\textbf {\bibinfo {volume} {111}},\ \bibinfo {pages}
  {032501} (\bibinfo {year} {2013})}\BibitemShut {NoStop}%
\bibitem [{\citenamefont {Epelbaum}\ \emph {et~al.}(2015)\citenamefont
  {Epelbaum}, \citenamefont {Krebs},\ and\ \citenamefont
  {Mei\ss{}ner}}]{Epelbaum2015Phys.Rev.Lett.122301}%
  \BibitemOpen
  \bibfield  {author} {\bibinfo {author} {\bibfnamefont {E.}~\bibnamefont
  {Epelbaum}}, \bibinfo {author} {\bibfnamefont {H.}~\bibnamefont {Krebs}}, \
  and\ \bibinfo {author} {\bibfnamefont {U.-G.}\ \bibnamefont {Mei\ss{}ner}},\
  }\href {\doibase 10.1103/PhysRevLett.115.122301} {\bibfield  {journal}
  {\bibinfo  {journal} {Phys. Rev. Lett.}\ }\textbf {\bibinfo {volume} {115}},\
  \bibinfo {pages} {122301} (\bibinfo {year} {2015})}\BibitemShut {NoStop}%
\bibitem [{\citenamefont {Piarulli}\ \emph {et~al.}(2018)\citenamefont
  {Piarulli}, \citenamefont {Baroni}, \citenamefont {Girlanda}, \citenamefont
  {Kievsky}, \citenamefont {Lovato}, \citenamefont {Lusk}, \citenamefont
  {Marcucci}, \citenamefont {Pieper}, \citenamefont {Schiavilla}, \citenamefont
  {Viviani},\ and\ \citenamefont {Wiringa}}]{Piarulli2018Phys.Rev.Lett.52503}%
  \BibitemOpen
  \bibfield  {author} {\bibinfo {author} {\bibfnamefont {M.}~\bibnamefont
  {Piarulli}}, \bibinfo {author} {\bibfnamefont {A.}~\bibnamefont {Baroni}},
  \bibinfo {author} {\bibfnamefont {L.}~\bibnamefont {Girlanda}}, \bibinfo
  {author} {\bibfnamefont {A.}~\bibnamefont {Kievsky}}, \bibinfo {author}
  {\bibfnamefont {A.}~\bibnamefont {Lovato}}, \bibinfo {author} {\bibfnamefont
  {E.}~\bibnamefont {Lusk}}, \bibinfo {author} {\bibfnamefont {L.~E.}\
  \bibnamefont {Marcucci}}, \bibinfo {author} {\bibfnamefont {S.~C.}\
  \bibnamefont {Pieper}}, \bibinfo {author} {\bibfnamefont {R.}~\bibnamefont
  {Schiavilla}}, \bibinfo {author} {\bibfnamefont {M.}~\bibnamefont {Viviani}},
  \ and\ \bibinfo {author} {\bibfnamefont {R.~B.}\ \bibnamefont {Wiringa}},\
  }\href {\doibase 10.1103/PhysRevLett.120.052503} {\bibfield  {journal}
  {\bibinfo  {journal} {Phys. Rev. Lett.}\ }\textbf {\bibinfo {volume} {120}},\
  \bibinfo {pages} {052503} (\bibinfo {year} {2018})}\BibitemShut {NoStop}%
\bibitem [{\citenamefont {Schiavilla}\ \emph {et~al.}(2021)\citenamefont
  {Schiavilla}, \citenamefont {Girlanda}, \citenamefont {Gnech}, \citenamefont
  {Kievsky}, \citenamefont {Lovato}, \citenamefont {Marcucci}, \citenamefont
  {Piarulli},\ and\ \citenamefont {Viviani}}]{Schiavilla2021Phys.Rev.C054003}%
  \BibitemOpen
  \bibfield  {author} {\bibinfo {author} {\bibfnamefont {R.}~\bibnamefont
  {Schiavilla}}, \bibinfo {author} {\bibfnamefont {L.}~\bibnamefont
  {Girlanda}}, \bibinfo {author} {\bibfnamefont {A.}~\bibnamefont {Gnech}},
  \bibinfo {author} {\bibfnamefont {A.}~\bibnamefont {Kievsky}}, \bibinfo
  {author} {\bibfnamefont {A.}~\bibnamefont {Lovato}}, \bibinfo {author}
  {\bibfnamefont {L.~E.}\ \bibnamefont {Marcucci}}, \bibinfo {author}
  {\bibfnamefont {M.}~\bibnamefont {Piarulli}}, \ and\ \bibinfo {author}
  {\bibfnamefont {M.}~\bibnamefont {Viviani}},\ }\href {\doibase
  10.1103/PhysRevC.103.054003} {\bibfield  {journal} {\bibinfo  {journal}
  {Phys. Rev. C}\ }\textbf {\bibinfo {volume} {103}},\ \bibinfo {pages}
  {054003} (\bibinfo {year} {2021})}\BibitemShut {NoStop}%
\bibitem [{\citenamefont {Bedaque}\ and\ \citenamefont {van
  Kolck}(2002)}]{Bedaque2002Ann.Rev.Nucl.Part.Sci.339}%
  \BibitemOpen
  \bibfield  {author} {\bibinfo {author} {\bibfnamefont {P.~F.}\ \bibnamefont
  {Bedaque}}\ and\ \bibinfo {author} {\bibfnamefont {U.}~\bibnamefont {van
  Kolck}},\ }\href {\doibase 10.1146/annurev.nucl.52.050102.090637} {\bibfield
  {journal} {\bibinfo  {journal} {Ann. Rev. Nucl. Part. Sci.}\ }\textbf
  {\bibinfo {volume} {52}},\ \bibinfo {pages} {339} (\bibinfo {year}
  {2002})}\BibitemShut {NoStop}%
\bibitem [{\citenamefont {Epelbaum}\ \emph {et~al.}(2009)\citenamefont
  {Epelbaum}, \citenamefont {Hammer},\ and\ \citenamefont
  {Mei\ss{}ner}}]{Epelbaum2009Rev.Mod.Phys.1773}%
  \BibitemOpen
  \bibfield  {author} {\bibinfo {author} {\bibfnamefont {E.}~\bibnamefont
  {Epelbaum}}, \bibinfo {author} {\bibfnamefont {H.-W.}\ \bibnamefont
  {Hammer}}, \ and\ \bibinfo {author} {\bibfnamefont {U.-G.}\ \bibnamefont
  {Mei\ss{}ner}},\ }\href {\doibase 10.1103/RevModPhys.81.1773} {\bibfield
  {journal} {\bibinfo  {journal} {Rev. Mod. Phys.}\ }\textbf {\bibinfo {volume}
  {81}},\ \bibinfo {pages} {1773} (\bibinfo {year} {2009})}\BibitemShut
  {NoStop}%
\bibitem [{\citenamefont {Machleidt}\ and\ \citenamefont
  {Entem}(2011)}]{Machleidt2011Phys.Rep.1}%
  \BibitemOpen
  \bibfield  {author} {\bibinfo {author} {\bibfnamefont {R.}~\bibnamefont
  {Machleidt}}\ and\ \bibinfo {author} {\bibfnamefont {D.}~\bibnamefont
  {Entem}},\ }\href {\doibase http://dx.doi.org/10.1016/j.physrep.2011.02.001}
  {\bibfield  {journal} {\bibinfo  {journal} {Phys. Rep.}\ }\textbf {\bibinfo
  {volume} {503}},\ \bibinfo {pages} {1} (\bibinfo {year} {2011})}\BibitemShut
  {NoStop}%
\bibitem [{\citenamefont {Chen}\ \emph {et~al.}(1999)\citenamefont {Chen},
  \citenamefont {Rupak},\ and\ \citenamefont
  {Savage}}]{Chen1999Nucl.Phys.A386}%
  \BibitemOpen
  \bibfield  {author} {\bibinfo {author} {\bibfnamefont {J.-W.}\ \bibnamefont
  {Chen}}, \bibinfo {author} {\bibfnamefont {G.}~\bibnamefont {Rupak}}, \ and\
  \bibinfo {author} {\bibfnamefont {M.~J.}\ \bibnamefont {Savage}},\ }\href
  {\doibase 10.1016/S0375-9474(99)00298-5} {\bibfield  {journal} {\bibinfo
  {journal} {Nucl. Phys. A}\ }\textbf {\bibinfo {volume} {653}},\ \bibinfo
  {pages} {386} (\bibinfo {year} {1999})}\BibitemShut {NoStop}%
\bibitem [{\citenamefont {Hammer}\ \emph {et~al.}(2020)\citenamefont {Hammer},
  \citenamefont {K\"onig},\ and\ \citenamefont {van
  Kolck}}]{Hammer2020Rev.Mod.Phys.025004}%
  \BibitemOpen
  \bibfield  {author} {\bibinfo {author} {\bibfnamefont {H.-W.}\ \bibnamefont
  {Hammer}}, \bibinfo {author} {\bibfnamefont {S.}~\bibnamefont {K\"onig}}, \
  and\ \bibinfo {author} {\bibfnamefont {U.}~\bibnamefont {van Kolck}},\ }\href
  {\doibase 10.1103/RevModPhys.92.025004} {\bibfield  {journal} {\bibinfo
  {journal} {Rev. Mod. Phys.}\ }\textbf {\bibinfo {volume} {92}},\ \bibinfo
  {pages} {025004} (\bibinfo {year} {2020})}\BibitemShut {NoStop}%
\bibitem [{\citenamefont {Yang}(2020)}]{Yang2020Eur.Phys.J.A96}%
  \BibitemOpen
  \bibfield  {author} {\bibinfo {author} {\bibfnamefont {C.-J.}\ \bibnamefont
  {Yang}},\ }\href {\doibase 10.1140/epja/s10050-020-00104-0} {\bibfield
  {journal} {\bibinfo  {journal} {Eur. Phys. J. A}\ }\textbf {\bibinfo {volume}
  {56}},\ \bibinfo {pages} {96} (\bibinfo {year} {2020})}\BibitemShut {NoStop}%
\bibitem [{\citenamefont {Kirscher}\ \emph {et~al.}(2015)\citenamefont
  {Kirscher}, \citenamefont {Barnea}, \citenamefont {Gazit}, \citenamefont
  {Pederiva},\ and\ \citenamefont {van Kolck}}]{Kirscher2015Phys.Rev.C054002}%
  \BibitemOpen
  \bibfield  {author} {\bibinfo {author} {\bibfnamefont {J.}~\bibnamefont
  {Kirscher}}, \bibinfo {author} {\bibfnamefont {N.}~\bibnamefont {Barnea}},
  \bibinfo {author} {\bibfnamefont {D.}~\bibnamefont {Gazit}}, \bibinfo
  {author} {\bibfnamefont {F.}~\bibnamefont {Pederiva}}, \ and\ \bibinfo
  {author} {\bibfnamefont {U.}~\bibnamefont {van Kolck}},\ }\href {\doibase
  10.1103/PhysRevC.92.054002} {\bibfield  {journal} {\bibinfo  {journal} {Phys.
  Rev. C}\ }\textbf {\bibinfo {volume} {92}},\ \bibinfo {pages} {054002}
  (\bibinfo {year} {2015})}\BibitemShut {NoStop}%
\bibitem [{\citenamefont {Contessi}\ \emph {et~al.}(2017)\citenamefont
  {Contessi}, \citenamefont {Lovato}, \citenamefont {Pederiva}, \citenamefont
  {Roggero}, \citenamefont {Kirscher},\ and\ \citenamefont {{van
  Kolck}}}]{Contessi2017Phys.Lett.B839}%
  \BibitemOpen
  \bibfield  {author} {\bibinfo {author} {\bibfnamefont {L.}~\bibnamefont
  {Contessi}}, \bibinfo {author} {\bibfnamefont {A.}~\bibnamefont {Lovato}},
  \bibinfo {author} {\bibfnamefont {F.}~\bibnamefont {Pederiva}}, \bibinfo
  {author} {\bibfnamefont {A.}~\bibnamefont {Roggero}}, \bibinfo {author}
  {\bibfnamefont {J.}~\bibnamefont {Kirscher}}, \ and\ \bibinfo {author}
  {\bibfnamefont {U.}~\bibnamefont {{van Kolck}}},\ }\href {\doibase
  10.1016/j.physletb.2017.07.048} {\bibfield  {journal} {\bibinfo  {journal}
  {Phys. Lett. B}\ }\textbf {\bibinfo {volume} {772}},\ \bibinfo {pages} {839}
  (\bibinfo {year} {2017})}\BibitemShut {NoStop}%
\bibitem [{\citenamefont {Phillips}(1968)}]{Phillips1968Nucl.Phys.A209216}%
  \BibitemOpen
  \bibfield  {author} {\bibinfo {author} {\bibfnamefont {A.}~\bibnamefont
  {Phillips}},\ }\href {\doibase https://doi.org/10.1016/0375-9474(68)90737-9}
  {\bibfield  {journal} {\bibinfo  {journal} {Nucl. Phys. A}\ }\textbf
  {\bibinfo {volume} {107}},\ \bibinfo {pages} {209} (\bibinfo {year}
  {1968})}\BibitemShut {NoStop}%
\bibitem [{\citenamefont {Tjon}(1975)}]{Tjon1975Phys.Lett.B217220}%
  \BibitemOpen
  \bibfield  {author} {\bibinfo {author} {\bibfnamefont {J.~A.}\ \bibnamefont
  {Tjon}},\ }\href {\doibase https://doi.org/10.1016/0370-2693(75)90378-0}
  {\bibfield  {journal} {\bibinfo  {journal} {Phys. Lett. B}\ }\textbf
  {\bibinfo {volume} {56}},\ \bibinfo {pages} {217} (\bibinfo {year}
  {1975})}\BibitemShut {NoStop}%
\bibitem [{\citenamefont {Adams}\ \emph {et~al.}(2021)\citenamefont {Adams},
  \citenamefont {Carleo}, \citenamefont {Lovato},\ and\ \citenamefont
  {Rocco}}]{Adams2021Phys.Rev.Lett.022502}%
  \BibitemOpen
  \bibfield  {author} {\bibinfo {author} {\bibfnamefont {C.}~\bibnamefont
  {Adams}}, \bibinfo {author} {\bibfnamefont {G.}~\bibnamefont {Carleo}},
  \bibinfo {author} {\bibfnamefont {A.}~\bibnamefont {Lovato}}, \ and\ \bibinfo
  {author} {\bibfnamefont {N.}~\bibnamefont {Rocco}},\ }\href {\doibase
  10.1103/PhysRevLett.127.022502} {\bibfield  {journal} {\bibinfo  {journal}
  {Phys. Rev. Lett.}\ }\textbf {\bibinfo {volume} {127}},\ \bibinfo {pages}
  {022502} (\bibinfo {year} {2021})}\BibitemShut {NoStop}%
\bibitem [{\citenamefont {Epelbaum}\ and\ \citenamefont
  {Gegelia}(2012)}]{Epelbaum2012Phys.Lett.B338}%
  \BibitemOpen
  \bibfield  {author} {\bibinfo {author} {\bibfnamefont {E.}~\bibnamefont
  {Epelbaum}}\ and\ \bibinfo {author} {\bibfnamefont {J.}~\bibnamefont
  {Gegelia}},\ }\href {\doibase https://doi.org/10.1016/j.physletb.2012.08.025}
  {\bibfield  {journal} {\bibinfo  {journal} {Phys. Lett. B}\ }\textbf
  {\bibinfo {volume} {716}},\ \bibinfo {pages} {338} (\bibinfo {year}
  {2012})}\BibitemShut {NoStop}%
\bibitem [{sup()}]{supp}%
  \BibitemOpen
  \href@noop {} {\ }\BibitemShut {NoStop}%
\bibitem [{\citenamefont {Carlson}\ and\ \citenamefont
  {Wiringa}(1991)}]{Carlson1991}%
  \BibitemOpen
  \bibfield  {author} {\bibinfo {author} {\bibfnamefont {J.~A.}\ \bibnamefont
  {Carlson}}\ and\ \bibinfo {author} {\bibfnamefont {R.~B.}\ \bibnamefont
  {Wiringa}},\ }\enquote {\bibinfo {title} {Variational monte carlo techniques
  in nuclear physics},}\ in\ \href {\doibase 10.1007/978-3-642-76356-4_9}
  {\emph {\bibinfo {booktitle} {Computational Nuclear Physics 1: Nuclear
  Structure}}},\ \bibinfo {editor} {edited by\ \bibinfo {editor} {\bibfnamefont
  {K.}~\bibnamefont {Langanke}}, \bibinfo {editor} {\bibfnamefont {J.~A.}\
  \bibnamefont {Maruhn}}, \ and\ \bibinfo {editor} {\bibfnamefont {S.~E.}\
  \bibnamefont {Koonin}}}\ (\bibinfo  {publisher} {Springer Berlin
  Heidelberg},\ \bibinfo {address} {Berlin, Heidelberg},\ \bibinfo {year}
  {1991})\ pp.\ \bibinfo {pages} {171--189}\BibitemShut {NoStop}%
\bibitem [{\citenamefont {Metropolis}\ \emph {et~al.}(1953)\citenamefont
  {Metropolis}, \citenamefont {Rosenbluth}, \citenamefont {Rosenbluth},
  \citenamefont {Teller},\ and\ \citenamefont
  {Teller}}]{Metropolis1953J.Chem.Phys.1087}%
  \BibitemOpen
  \bibfield  {author} {\bibinfo {author} {\bibfnamefont {N.}~\bibnamefont
  {Metropolis}}, \bibinfo {author} {\bibfnamefont {A.~W.}\ \bibnamefont
  {Rosenbluth}}, \bibinfo {author} {\bibfnamefont {M.~N.}\ \bibnamefont
  {Rosenbluth}}, \bibinfo {author} {\bibfnamefont {A.~H.}\ \bibnamefont
  {Teller}}, \ and\ \bibinfo {author} {\bibfnamefont {E.}~\bibnamefont
  {Teller}},\ }\href {\doibase 10.1063/1.1699114} {\bibfield  {journal}
  {\bibinfo  {journal} {The Journal of Chemical Physics}\ }\textbf {\bibinfo
  {volume} {21}},\ \bibinfo {pages} {1087} (\bibinfo {year}
  {1953})}\BibitemShut {NoStop}%
\bibitem [{\citenamefont {Lomnitz-Adler}\ \emph {et~al.}(1981)\citenamefont
  {Lomnitz-Adler}, \citenamefont {Pandharipande},\ and\ \citenamefont
  {Smith}}]{LomnitzAdler1981Nucl.Phys.A399}%
  \BibitemOpen
  \bibfield  {author} {\bibinfo {author} {\bibfnamefont {J.}~\bibnamefont
  {Lomnitz-Adler}}, \bibinfo {author} {\bibfnamefont {V.}~\bibnamefont
  {Pandharipande}}, \ and\ \bibinfo {author} {\bibfnamefont {R.}~\bibnamefont
  {Smith}},\ }\href {\doibase 10.1007/978-3-642-76356-4} {\bibfield  {journal}
  {\bibinfo  {journal} {Nucl. Phys. A}\ }\textbf {\bibinfo {volume} {361}},\
  \bibinfo {pages} {399 } (\bibinfo {year} {1981})}\BibitemShut {NoStop}%
\bibitem [{\citenamefont {Dugas}\ \emph {et~al.}(2001)\citenamefont {Dugas},
  \citenamefont {Bengio}, \citenamefont {B\'elisle}, \citenamefont {Nadeau},\
  and\ \citenamefont {Garcia}}]{C.Dugas2001472478}%
  \BibitemOpen
  \bibfield  {author} {\bibinfo {author} {\bibfnamefont {C.}~\bibnamefont
  {Dugas}}, \bibinfo {author} {\bibfnamefont {Y.}~\bibnamefont {Bengio}},
  \bibinfo {author} {\bibfnamefont {F.}~\bibnamefont {B\'elisle}}, \bibinfo
  {author} {\bibfnamefont {C.}~\bibnamefont {Nadeau}}, \ and\ \bibinfo {author}
  {\bibfnamefont {R.}~\bibnamefont {Garcia}},\ }\enquote {\bibinfo {title}
  {Incorporating second-order functional knowledge for betteroption pricing},}\
  \ (\bibinfo  {publisher} {MIT Press. Cambridge, MA, 2001},\ \bibinfo {year}
  {2001})\ Chap.\ \bibinfo {chapter} {Advances in Neural Information Processing
  Systems 13}, pp.\ \bibinfo {pages} {472--478}\BibitemShut {NoStop}%
\bibitem [{\citenamefont {Sorella}(1998)}]{Sorella1998Phys.Rev.Lett.45584561}%
  \BibitemOpen
  \bibfield  {author} {\bibinfo {author} {\bibfnamefont {S.}~\bibnamefont
  {Sorella}},\ }\href {\doibase 10.1103/PhysRevLett.80.4558} {\bibfield
  {journal} {\bibinfo  {journal} {Phys. Rev. Lett.}\ }\textbf {\bibinfo
  {volume} {80}},\ \bibinfo {pages} {4558} (\bibinfo {year}
  {1998})}\BibitemShut {NoStop}%
\bibitem [{\citenamefont {Sorella}(2005)}]{Sorella2005Phys.Rev.B241103}%
  \BibitemOpen
  \bibfield  {author} {\bibinfo {author} {\bibfnamefont {S.}~\bibnamefont
  {Sorella}},\ }\href {\doibase 10.1103/PhysRevB.71.241103} {\bibfield
  {journal} {\bibinfo  {journal} {Phys. Rev. B}\ }\textbf {\bibinfo {volume}
  {71}},\ \bibinfo {pages} {241103} (\bibinfo {year} {2005})}\BibitemShut
  {NoStop}%
\bibitem [{\citenamefont {Thomas}(1935)}]{Thomas1935Phys.Rev.903}%
  \BibitemOpen
  \bibfield  {author} {\bibinfo {author} {\bibfnamefont {L.~H.}\ \bibnamefont
  {Thomas}},\ }\href {\doibase 10.1103/PhysRev.47.903} {\bibfield  {journal}
  {\bibinfo  {journal} {Phys. Rev.}\ }\textbf {\bibinfo {volume} {47}},\
  \bibinfo {pages} {903} (\bibinfo {year} {1935})}\BibitemShut {NoStop}%
\bibitem [{\citenamefont {Bedaque}\ \emph {et~al.}(1999)\citenamefont
  {Bedaque}, \citenamefont {Hammer},\ and\ \citenamefont {van
  Kolck}}]{Bedaque1999Phys.Rev.Lett.463}%
  \BibitemOpen
  \bibfield  {author} {\bibinfo {author} {\bibfnamefont {P.~F.}\ \bibnamefont
  {Bedaque}}, \bibinfo {author} {\bibfnamefont {H.-W.}\ \bibnamefont {Hammer}},
  \ and\ \bibinfo {author} {\bibfnamefont {U.}~\bibnamefont {van Kolck}},\
  }\href {\doibase 10.1103/PhysRevLett.82.463} {\bibfield  {journal} {\bibinfo
  {journal} {Phys. Rev. Lett.}\ }\textbf {\bibinfo {volume} {82}},\ \bibinfo
  {pages} {463} (\bibinfo {year} {1999})}\BibitemShut {NoStop}%
\bibitem [{\citenamefont {Bazak}\ \emph {et~al.}(2019)\citenamefont {Bazak},
  \citenamefont {Kirscher}, \citenamefont {K\"onig}, \citenamefont
  {Valderrama}, \citenamefont {Barnea},\ and\ \citenamefont {van
  Kolck}}]{Bazak2019Phys.Rev.Lett.143001}%
  \BibitemOpen
  \bibfield  {author} {\bibinfo {author} {\bibfnamefont {B.}~\bibnamefont
  {Bazak}}, \bibinfo {author} {\bibfnamefont {J.}~\bibnamefont {Kirscher}},
  \bibinfo {author} {\bibfnamefont {S.}~\bibnamefont {K\"onig}}, \bibinfo
  {author} {\bibfnamefont {M.~P.}\ \bibnamefont {Valderrama}}, \bibinfo
  {author} {\bibfnamefont {N.}~\bibnamefont {Barnea}}, \ and\ \bibinfo {author}
  {\bibfnamefont {U.}~\bibnamefont {van Kolck}},\ }\href {\doibase
  10.1103/PhysRevLett.122.143001} {\bibfield  {journal} {\bibinfo  {journal}
  {Phys. Rev. Lett.}\ }\textbf {\bibinfo {volume} {122}},\ \bibinfo {pages}
  {143001} (\bibinfo {year} {2019})}\BibitemShut {NoStop}%
\bibitem [{\citenamefont {Hellmann}(1937)}]{Hellmann1937}%
  \BibitemOpen
  \bibfield  {author} {\bibinfo {author} {\bibfnamefont {H.}~\bibnamefont
  {Hellmann}},\ }\href@noop {} {\emph {\bibinfo {title} {Einf\"uhrung in die
  Quantenchemie}}}\ (\bibinfo  {publisher} {Deuticke, Wien},\ \bibinfo {year}
  {1937})\BibitemShut {NoStop}%
\bibitem [{\citenamefont {Feynman}(1939)}]{Feynman1939Phys.Rev.340}%
  \BibitemOpen
  \bibfield  {author} {\bibinfo {author} {\bibfnamefont {R.~P.}\ \bibnamefont
  {Feynman}},\ }\href {\doibase 10.1103/PhysRev.56.340} {\bibfield  {journal}
  {\bibinfo  {journal} {Phys. Rev.}\ }\textbf {\bibinfo {volume} {56}},\
  \bibinfo {pages} {340} (\bibinfo {year} {1939})}\BibitemShut {NoStop}%
\end{thebibliography}

%

\end{document}